\documentclass[twocolumn]{article}

\usepackage[utf8]{inputenc}
\usepackage{geometry}
\usepackage{graphicx}
\usepackage{amsmath}
\usepackage{amssymb}
\usepackage{booktabs}      
\usepackage{authblk}       
\usepackage{hyperref}      
\usepackage{caption}       

\title{Large Language Model–Derived Priors Can Improve Bayesian Survival Analyses: A Glioblastoma Application}

\author[1,*]{Richard Evans, PhD}
\author[1]{Max Felland, BA}
\author[2]{Susanna Evans, BA}
\author[1]{Lindsey Sloan, MD, PhD}

\affil[1]{University of Minnesota, Minneapolis, MN, USA}
\affil[2]{Medical College of Wisconsin, Milwaukee, WI, USA}
\affil[*]{Corresponding author: evan0770@umn.edu}

\date{} 

\newcommand{\keywords}[1]{%
    \small
    \noindent\textbf{Keywords:} #1
}

\begin{document}

\maketitle

\begin{abstract}
This report describes an application of artificial intelligence (AI) to the Bayesian analysis of glioblastoma survival data. It has been suggested that AI can be used to construct prior distributions for parameters in Bayesian models rather than using the difficult, unreliable, and time-consuming process of eliciting expert opinion from radiation oncologists. Here, we show how generative AI can quickly propose sensible prior distributions of the hazard ratio comparing two glioblastoma therapies, for a standard Bayesian survival model on real data. Three Chatbots generated two alternative priors each which were evaluated by a radiation oncologist and then used in a sensitivity analysis to assess posterior stability. The results suggest that, for this cancer survival analysis, priors from generative AI are a preferred alternative method to expert elicitation. 
\end{abstract}

\noindent
\small
\textbf{CCS CONCEPTS}
\begin{itemize}
    \item \textbf{Computing Methodologies}
    \item \textbf{Artificial Intelligence}
    \item \textbf{Natural language processing}
\end{itemize}


\section{Introduction}
Glioblastoma (GBM) is the most aggressive and common primary malignant brain cancer in adults. Even with medical therapy, only six percent of patients survive to five years. US Senator and former presidential candidate John McCain lived for 13 months post diagnosis, which is consistent with the 12- to 15-month median survival time. It is aggressive and hard to treat.

GBMs infiltrate normal brain tissue, precluding surgery, so that radiation and chemotherapy are the primary interventions. Radiation oncologists who treat GBM face an additional challenge—it is not a common cancer overall, so there are relatively few patients available for clinical trials. That means clinical studies are often underpowered in the classical statistical decision-making sense, so that only large effect sizes achieve statistical significance. Whether correct or not, it is common in the medical literature to disregard results with $P > 0.05$, regardless of the effect size, limiting doctors’ GBM-treating toolkit.

A solution to the small sample size problem is a Bayesian analysis, which is different from the frequentist analysis in two ways. First, it asks a different question. The frequentist approach asks, \textit{under the null hypothesis of no treatment effects, what is the probability of observing a test statistic at least as extreme as the one observed?} In contrast, the Bayesian analysis asks, \textit{what is the probability that one treatment is better than another?} Second, the Bayesian analysis can readily incorporate information from sources other than the observed data. One common source of additional information is expert opinion, which, for example, might be elicited from a radiation oncologist who treats GBM.

However, the challenges of eliciting sensible expert opinion from humans is well documented. There are three categories of reasons that make it difficult to elicit expert opinion. First, humans are often overconfident, overemphasize memorable research or patients, and are anchored in their initial training. Second, they find it hard to translate their knowledge into density parameters. Third, elicitation can be exceptionally time consuming, especially from medical experts, who apart from crowded schedules, need coaching in the language of probabilities and densities.

Recently, there has been promising research in using large language models (LLMs) to use data to generate prior distributions. Selby et al. (2024) \cite{selby2024} evaluated the feasibility of eliciting priors directly from LLMs, showing that their “expert-like” priors can approximate human elicitation, but also exhibit overconfidence and bias. Arai et al. (2025) \cite{arai2025} demonstrated that LLM-derived priors in hierarchical Bayesian models for adverse event modeling in clinical trials can reduce required sample sizes while maintaining predictive accuracy, perhaps making trials more efficient. Gouk \& Gao (2024) \cite{gouk2024} proposed an automated system for Bayesian logistic regression that elicits priors by having an LLM generate synthetic data. Huang (2025) \cite{huang2025} introduced the \texttt{LLMPrior} framework, which couples an LLM with explicit probabilistic generators (e.g., Gaussian Mixture Models) to ensure mathematically valid priors. Finally, Riegler et al. (2025) \cite{riegler2025} empirically tested ChatGPT-4o, Gemini 2.5, and Claude Opus having them suggest priors for real regression datasets, finding that all captured correct association directions, with Claude yielding balanced priors, while others alternated between over- and under-confidence.

We follow this last approach but for a survival model on a small dataset. The main advantages of using LLMs is the speed of ``elicitation'', minutes rather than weeks, and the ability of LLMs to (a) summarize medical research (so we know what information they are using) and then (b) to quickly combine medical research results and finally (c) provide them directly in terms of probabilistic objects, such as parametric densities. To say it another way, a LLM can take medical information (e.g., clinical studies) and immediately provide justifiable parameters for a Weibull distribution. A radiation oncologist cannot.

The objective of this analysis is to demonstrate the application of LLMs to a Bayesian survival analysis comparing two therapies using real GBM data, and suggest this approach is both an efficient alternative to eliciting information from subject matter experts and provides a more useful method than frequentist analysis of small survival datasets.

\section{The Data}
The data are 28 adult glioblastoma patients. Eight patients underwent conventional radiation therapy (CRT), and twenty underwent hypofractionated radiation therapy (HFRT). CRT is considered the reference therapy. The dataset has three variables: survival time in months (the time to event), a binary event variable, where 1 is death and 0 is a censored observation, and the grouping variable which has values CRT or HFRT.

\section{The Model}
The model was a standard Bayesian Cox model. The setup is
$$ z_i \in \{0, 1\} \text{ indicates group } (0=\text{CRT, } 1=\text{HFRT}) $$
$$ t_i \text{ is the survival time and } \delta_i \in \{0, 1\} \text{ is the event indicator.} $$
The hazard is
$$ h_i(t) = h_0(t)\exp(\beta z_i). $$
The hazard ration is $\text{HR} = \exp(\beta)$. Using those equations the partial likelihood is,
$$ R(t_i) = \text{subjects at risk just before } t_i. $$
$$ L(\beta) = \prod_{i=1}^{n} \left[ \frac{\exp(\beta z_i)}{\sum_{j \in R(t_i)} \exp(\beta z_j)} \right]^{\delta_i} $$
Finally, the posterior distribution of $\beta$ is,
\begin{equation}
    p(\beta | D) \propto L(\beta) p(\beta),
\end{equation}
where $D$ is the data. Equation (1) was then evaluated using R (4.4.1) with \texttt{brms} (2.23.0), which fits Bayesian multi-level models using \texttt{stan} (2.37) on the backend.

\section{The Priors}
Equation (1) is of primary interest because that is where the LLM-generated priors enter the model. We set
$$ \beta \sim N(\mu, \sigma^2) $$
because frequentist estimates of $\beta$ are asymptotically normal. Finally,
$$ \exp(\beta) \sim \text{LogNormal}(\mu, \sigma^2) $$
The term $\text{LogNormal}(\mu, \sigma^2)$ is shorthand for: “A random variable whose logarithm is $\text{Normal}(\mu, \sigma^2)$.”

The Chatbot responses are summarized in Table \ref{tab:priors}.

\begin{table*}[t]
    \centering
    \caption{AI-Generated Priors for the HR}
    \label{tab:priors}
    \begin{tabular}{@{}llll@{}}
        \toprule
        \textbf{Source} & \textbf{Prior Type} & \textbf{LogNormal (HR scale)} & \textbf{Notes} \\
        \midrule
        ChatGPT & Informative & HR $\sim$ LogNormal($\mu = 0.431$, $\sigma = 0.30$) & Median HR $\approx$ 1.54, 95\% HR $\approx$ (0.85, 2.77) \\
                & Non-informative & HR $\sim$ LogNormal($\mu = 0$, $\sigma = 1$) & Median HR = 1.0, 95\% HR $\approx$ (0.14, 7.1) \\
        \addlinespace
        Gemini  & Informative & HR $\sim$ LogNormal($\mu = 0.095$, $\sigma = 0.18$) & Median HR $\approx$ 1.10, 95\% HR $\approx$ (0.77, 1.57) \\
                & Non-informative & HR $\sim$ LogNormal($\mu = 0$, $\sigma = 2$) & Median HR = 1.0, 95\% HR $\approx$ (0.02, 50.4) \\
        \addlinespace
        Grok    & Informative & HR $\sim$ LogNormal($\mu = 0.068$, $\sigma = 0.093$) & Median HR $\approx$ 1.07, 95\% HR $\approx$ (0.89, 1.28) \\
                & Non-informative & HR $\sim$ LogNormal($\mu = 0$, $\sigma = 31.62$) & Median HR = 1.0, diffuse \\
        \bottomrule
    \end{tabular}
\end{table*}

\begin{figure}[h]
    \centering
    \includegraphics[width=\columnwidth]{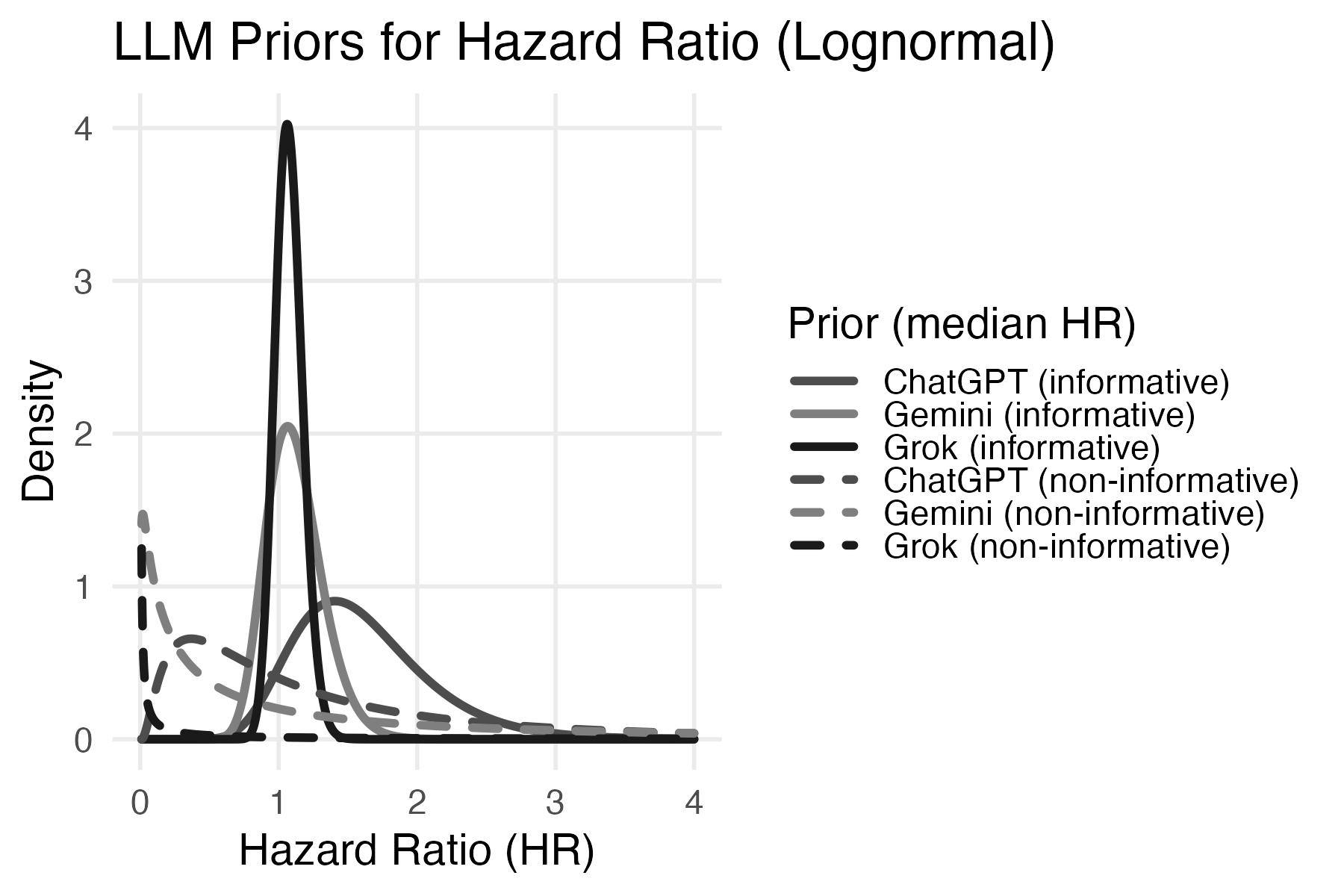}
    \caption{LLM prior distributions for the HR. Blue curves are the informative priors, Reddish curves are the non-informative priors.}
    \label{fig:priors}
\end{figure}

\subsection{The Chatbot prompt}
{\ttfamily  
    You are a radiation oncologist assisting a biostatistician in the Bayesian analysis of a Cox survival model of glioblastoma data from an adult population.
    The Cox model compares two groups using the hazard ratio.
    \begin{enumerate}
        \item The name of the treatment group is called ``group.'' There are two groups, called HFRT and CRT. HFRT is a treatment with the radiation fractionation schedule having fewer, larger fractions. CRT is a treatment standard using radiation.
        \item The variable ``time\_os\_months'' is the survival time in months.
        \item The binary (0/1) event variable is called ``event'', with 1 = dead and 0 = censored.
    \end{enumerate}
    You, the radiation oncologist, has to develop the prior distribution for a hazard ratio of HFRT and CRT, where CRT is the reference group, using glioblastoma literature.

    The response should include:
    \begin{enumerate}
        \item The results of review of the information on HFRT and CRT trials in the glioblastoma literature.
        \item An informative log-normal prior for the hazard ratio (which is on the log scale)
        \item A justification of the informative prior using the information on HFRT and CRT trials in the glioblastoma literature.
        \item A non-informative log-normal prior for the hazard ratio (which is on the log scale), to compare to the informative one.
    \end{enumerate}
} 

\section{Results}
The posterior probabilities and HR estimates are shown in Table \ref{tab:results}.

\begin{table}[h]
    \centering
    \footnotesize 
    \caption{Posterior Probabilities and HR estimates. The 2.5\% and 97.5\% columns represent the 95\% credible interval.} 
    \label{tab:results}
    \begin{tabular}{@{}lllll@{}}
        \toprule
        \textbf{Prior} & $\mathbf{Pr(HR > 1)}$ & \textbf{Median HR} & \textbf{2.5\%} & \textbf{97.5\%} \\
        \midrule
        Grok non-info & 0.974 & 2.761 & 0.989 & 10.396 \\
        ChatGPT       & 0.977 & 2.740 & 1.016 & 9.605 \\
        Gemini        & 0.975 & 2.680 & 1.001 & 9.410 \\
        Grok          & 0.975 & 2.710 & 0.997 & 9.366 \\
        \bottomrule
    \end{tabular}
\end{table}

The key feature of Table \ref{tab:results} is $P(\text{HR} > 1)$. This is the probability that the hazard ratio is greater than one, (i.e., the probability that HFRT has a larger hazard than CRT). Note the probabilities are quite convincing that HFRT is inferior to CRT, and the probabilities under all the LLMs are similar. That means that our inferences about HFRT vs CRT would be the same under any of the LLM priors. The informative priors (the last three rows of the Table \ref{tab:results}) give the essentially the same answers as Grok’s non-informative prior, which mean they were fairly conservative. The point estimates and the confidence intervals are also all about the same.

\section{Discussion}
This report presented a data analysis that used prior information, in the form of log-normal distributions, generated by AI. It was considerably easier to obtain priors from LLMs than from humans, saving weeks of research time. The LLMs based their priors on historical data. All three were conservative, sensible as judged by a radiation oncologist, and produced nearly equivalent estimates of hazard ratios and inferences.

An important point to note is that we should judge the LLMs using human-generated priors as a reference, not by how they created their priors, nor even if they are reproducible. It would be unlikely that three human experts would agree to the extent that ChatGPT, Gemini, and Grok agreed.

The obvious question is, how do we know the LLM priors are meaningful? There are four responses to that question. First, we know that human subject matter experts are not ``correct,'' because different experts often give very different elicitations. Second, no model is correct in the sense that often many models can represent a system with about the same accuracy. Third, good medical decisions are based on multiple studies produced by different teams, not one study. A study with LLM priors should be one of those studies, not the final decision point. Finally, the LLM information can undergo a “sanity check” with a medical expert, as well as a standard Bayesian analysis check for result robustness. In other words, we reverse engineer the prior information. Instead of eliciting it from radiation oncologists and then validating it, we provide the radiation oncologists with prior information in the form of distributions and have them validate it.


\end{document}